# A 3D Finite Element evaluation of the exophthalmia reduction

Vincent Luboz[1], Annaig Pedrono[2], Frank Boutault[3], Pascal Swider[2], Yohan Payan[1]

1. Laboratoire TIMC-GMCAO, UMR CNRS 5525, Faculté de Médecine Domaine de la Merci, 38706 La Tronche, France
2. INSERM, Biomécanique, CHU Purpan, BP 3103, 31026 Toulouse Cedex 3, France      3. CHU Purpan, 31026 Toulouse Cedex 3, France

## Introduction

The exophthalmia is a pathology characterised by an excessive forward protrusion of the ocular globe outside the orbit (Figure 1 (a)) (see [1] for the anatomy of the orbit). Its consequences are functional as well as aesthetical. Exophthalmia can be due to a trauma, a tumour, an infection, or can be associated to disthyroidy. Disthyroidy is characterised by a dysfunction of the thyroid that involves an increase of the soft tissues volume (muscles and fat). Once the endocrinal situation is stabilized, surgery may be needed. A classical surgical technique consists in decompressing the orbit [2] [3] by opening the walls, and pushing the ocular globe in order to evacuate some of the fat tissues inside the sinuses (Figure 1 (b)). This gesture is very delicate and could be assisted in a computer-aided framework. From this point of view, two aspects could be addressed. First, a computer-guided system could help the surgeon in localizing the tip of the surgical tool inside the orbit. Second, a computer-based modelling environment could assist the surgeon in the planning definition: where opening the orbit walls and to which extend? These questions are directly related to the surgical objective that is expressed in term of suited backward displacement of the ocular globe. This paper addresses the second point, by proposing a biomechanical 3D Finite Element (FE) model of the orbit, including the fat tissues and the walls. First qualitative simulations of walls opening are provided and compared with observed phenomena.

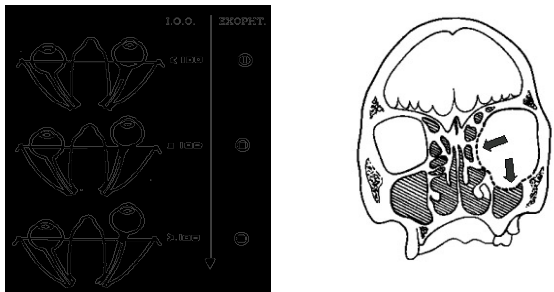

*Figure 1: (a) at the left; three degrees of exophthalmia for the right eye (from top to bottom, the protrusion becomes more and more severe). (b) At the right; decompression in the sinuses of the right eye, the fat tissues will be evacuated inside those new cavities.*

## Finite Element modelling of the orbital decompression

In order to build a Finite Element model of the intra-orbital soft tissues, tomodensitometry (TDM) images of the orbital cavity are first segmented to extract the bone limits, the muscles and the ocular nerve contours. The segmentation process is done manually since the bone limit of the orbit and the muscles do not appear clearly on CT exams. The procedure used is based on splines defined by few control points (never more than ten points). On each CT slice; a spline is created to segment the contour of the orbit. This segmentation extracts fifteen to twenty five splines (one for each slice containing the orbit) for the orbital cavity. The same process is applied to the muscles and the nerve leading to sets of splines representing those structures. From these extractions, the structures of the muscles, the nerve and the orbit can be extrapolated from the splines. A first quantitative evaluation of the soft and fat tissues volume can therefore be measured and provided to the surgeon.

Besides the volume evaluation of the intra orbital soft tissues, the surgery planning has been studied by the way of a 3D Finite Element mesh. Since the decompression act is a relatively rare surgery, the orbit has not been modelled by a FE mesh previously. Nevertheless, the ocular globe has been studied using a FE model [4]. Our FE orbital cavity model has been manually meshed with hexahedrons and wedges to accurately fit the splines obtained after the segmentation step. The mesh obtained is complex since it has to respect the topography of the orbit, that is to say it contains above 3000 elements (Figure 2).

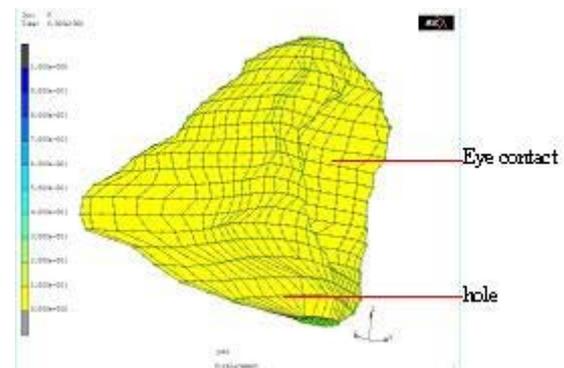

*Figure 2: result of the manually meshing step of the orbit, based on the set of segmented splines. All of the Finite Elements are hexahedrons or wedges. The hole is located at the bottom of the mesh, above the maxillary sinus.*

By modelling those intra orbital soft tissues with a poro-elastic FE material (with the MARC™ Finite Element software), a simulation of the globe backward displacement was computed. In order to simulate a wall opening, a hole was made inside the orbital cavity, approximately above the maxillary sinus. As a consequence of a backward pressure applied to the globe, fat tissues displacements through this hole are thus provided by the FE simulation. The elastic and the fluid properties simulate the behaviour of the intra orbital tissues. This allows a quantitative evaluation of the link between the backward movement of the globe and the size and location of the wall opening. In this study, the FE model was supposed to be composed by an only isotropic material modelling the fat tissues. The Young modulus and the Poisson ratio were respectively set to 20kPa and to 0.45 (according to Fung studies on human soft tissues properties [5]) and the permeability parameter was supposed to be equal to 100mm$^4$/N.s.

For instance, only one form of hole has been studied. It has a surface of 2.3cm² (Figure 3). The displacement induced through this hole and the behaviour of its elements were interesting. That is to say, by applying a displacement of 3mm to the ocular globe, the induced effort applied to the globe was evaluated to 15N and the elements of the hole moved of 4.1mm (for the maximum displacement). Those results seem to be realistic but need to be confirmed by future clinical measurements.

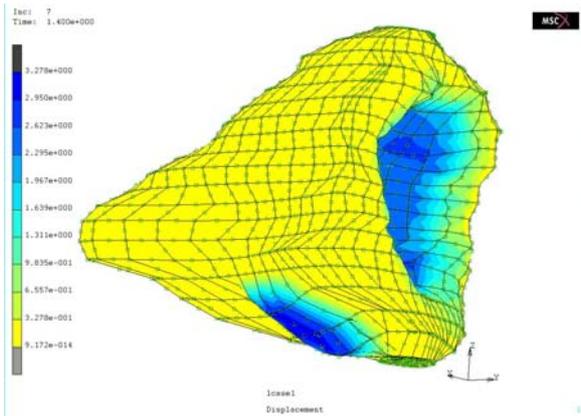

*Figure 3: behaviour of the orbit mesh. Under a displacement of 3mm, the elements of the hole move of 4.1mm while the elements near the ocular globe move backward.*

Though those results are encouraging for the prediction of the position of the hole and its size, the time required for a FE simulation with our model does not fulfil the needs of the surgeon. Indeed, on a PC equipped with a 1GHz processor and 1Go of memory, the process lasts nearly 5 hours which is far from the real-time model needed for a computer assisted surgery.

To deal with this constraint, a simple analytic model of the orbit was used. This model was reduced to a sphere lying on a cone (see Figure 4 for the dimension).

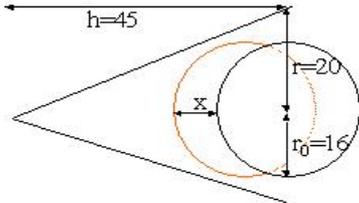

*Figure 4: the analytical model of the orbit and its dimensions.*

With this analytical model, the exophthalmia reduction volume $\Delta V$ can be computed immediately with the equation :

$$\Delta V = \frac{\pi}{3}(\frac{r}{h})^2(x^3 - 3x^2h + 3h^2x)$$

To verify whether this analytical model can be applied instead of the FE model for the computation of the volume, the two models have been compared for given displacements of the globe (Figure 5).

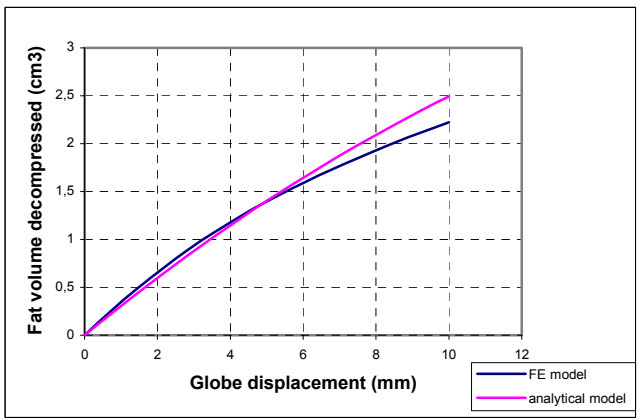

*Figure 5: volume computed for a given displacement by the FE model (in blue) and by the analytical model (in red).*

It is interesting to note that both curves have the same shape and moreover, from 0 to 4mm, both curves are nearly linear. This is noticeable since a backward displacement during the surgery should be in this interval. Consequently, the volume of fat tissues affected by the decompression can be computed quickly by this analytical model.

Though this analytical model is efficient for the volume estimation during the planning; the FE model remained necessary for the surgeon since it can help to evaluate the size and the location of the hole for the exophthalmia reduction. Therefore, the two models are currently used for the planning and future experiments on a large number of patients will be necessary to tell if one of the two can be leaved or not.

**Conclusion**

This paper has presented a first evaluation of the feasibility of Finite Element modelling of the orbital decompression, in the context of exophthalmia. First simulations were carried out with data extracted from a patient TDM exam. Results seem to qualitatively validate the feasibility of the simulations, with a Finite Element analysis that converges and provides a backward movement of the ocular globe associated with displacements of the fat tissues through the sinuses. This FE model can helped a surgeon for the planning of the exophthalmia reduction, and especially for the position and the size of the decompression hole. To get an estimation of the fat tissues volume affected by the surgery, the analytical model seems to provide quicker results for an equivalent efficiency.

Future works will focus 1) on rheological analyses of the human soft tissues to get better FE properties and 2) on the evaluation of the FE model on other patients. This study will allow to evaluate the behaviour of the model in spite of the differences between each patient and thus will determine if the complex FE mesh of the orbit could be assimilate to a FE meshed cone (similar to the analytical model) or not. If it is not possible, the meshing step complexity could be simplified by adapting a reference orbit mesh to each patient orbit [6].